\documentclass[conference,a4paper]{IEEEtran}

\IEEEoverridecommandlockouts

\usepackage{graphicx} 
\usepackage{epstopdf} 
\usepackage{cite}
\usepackage{amsmath,amssymb,amsfonts}
\usepackage{algorithmic}
\usepackage{graphicx}
\usepackage{textcomp}
\usepackage{xcolor}
\usepackage{color}

\usepackage{amssymb}
\usepackage{bbm} 
\usepackage{bm}

\usepackage{algorithm}
\usepackage{algorithmic}
\usepackage{url}
\usepackage{color}
\usepackage{bbding}

\usepackage{graphicx} 
\usepackage{epstopdf} 

\usepackage{cite}
\usepackage{amsmath,amssymb,amsfonts}
\usepackage{algorithmic}
\usepackage{graphicx}
\usepackage{textcomp}
\usepackage{xcolor}
\usepackage{color}

\usepackage{amssymb}
\usepackage{bbm} 
\usepackage{bm}

\usepackage{algorithm}
\usepackage{algorithmic}
\usepackage{url}
\usepackage{color}

\usepackage{multirow} 

\usepackage{diagbox}

\usepackage{subfig}

\usepackage{tablefootnote}

\usepackage{caption}

\usepackage{amsmath}
\allowdisplaybreaks[1]

\usepackage{booktabs}
\usepackage{threeparttable}
\usepackage{footnote}
\usepackage{enumerate}

\usepackage{array}

\usepackage{geometry}
\begin{document}

\title{FoV Privacy-aware VR Streaming}
\author{\IEEEauthorblockN{Xing Wei and Chenyang Yang
}
\IEEEauthorblockA{School of Electronics and Information Engineering, Beihang University, Beijing 100191, China\\ Email: \{weixing, cyyang\}@buaa.edu.cn}
}

\maketitle

\begin{abstract}
    Proactive tile-based virtual reality (VR) video streaming can use the trace of FoV and eye movement to predict future requested tiles, then renders and delivers the predicted tiles before playback. The  quality of experience (QoE) depends on the combined effect of tile prediction and consumed resources.
    Recently, it has been found that with the FoV and eye movement data collected for a user, one can infer the identity and preference of the user. Existing works investigate the privacy protection for eye movement, but never address how to protect the privacy in terms of FoV and how the privacy protection affects the QoE. In this paper, we strive to characterize and satisfy the FoV privacy requirement. We consider ``\textit{trading resources for privacy}". We first add \textit{camouflaged} tile requests around the real FoV and define spatial degree of privacy (SDoP) as a normalized number of camouflaged tile requests. By consuming more resources to ensure SDoP, the real FoVs can be hidden. 
    Then, we proceed to analyze the impacts of SDoP on the QoE by jointly optimizing the durations for prediction, computing, and transmission that maximizes the QoE given arbitrary predictor, configured resources, and SDoP. We find that a larger SDoP requires more resources but degrades the performance of tile prediction. 
    Simulation with state-of-the-art predictors on a real dataset verifies the analysis and shows that a user requiring a larger SDoP can be served with better QoE.
\end{abstract}

\vspace{-1mm}
\section{Introduction}
\vspace{-1mm}
As the main type of wireless virtual reality (VR) services, 360$^\circ$ video streaming consumes large amount of computing and communication resources to avoid playback stalls and black holes that degrade the quality of experience (QoE).
However, humans can only see a small portion of the full panoramic view (e.g., $110^\circ\times110^\circ$) at arbitrary time, which is called the field of view (FoV).
To improve QoE with limited resources, proactive VR video streaming is proposed \cite{optimizing_VR}, which divides a full panoramic view segment into small tiles in spatial domain. Before playback of the segment, the tiles overlapped with FoV are first predicted using the observed trace of FoV and eye movement, then these tiles are rendered and delivered to the user. The QoE depends on how many correctly predicted tiles can be rendered and transmitted, i.e., the combined effect of tile prediction and consumed resources. Harnessing the trace of FoV and eye movement is the key for prediction.

While proactive VR video streaming is being intensively investigated, most of the existing works neglect the willingness of users. Are users willing to share their traces of FoV and eye movement when watching 360$^\circ$ videos?


Recent works show that both types of data can be used to dig personal information. With less than five minutes' FoV trace, the identity of 95\% users among all the 511 users are correctly identified by
a random forest algorithm in \cite{privacy_VR_identifiability}. With the eye movement trace, a statistical approach proposed in \cite{eye_indentify_user} can achieve an equal error rate of 2.04\% to identify users. Furthermore, the FoV and eye movement data of a user collected for proactive VR video streaming can reveal the intent and preference of the user \cite{privacy-preserving_eye_tracking_2021}. Undoubtedly, not all users are willing to share these data.

Existing works have only considered to protect the privacy of eye movement either by adding noise  \cite{privacy-preserving_eye_tracking_2021,privacy_def_eye_track} or down-sampling the real gaze position \cite{privacy-preserving_eye_tracking_2021}. However, the trace of FoV plays a more important role for tile requests prediction. For example, the state-of-the-art prediction accuracy is achieved on a real dataset \cite{NTHU_dataset} only with the trace of FoV \cite{TRACK}.
When considering FoV privacy protection in proactive VR video streaming, several fundamental questions arise:
\textit{How to characterize and satisfy the FoV privacy requirement? What is the impact of privacy requirements on proactive tile-based VR video streaming}?
In this paper, we strive to answer these questions. Our contributions are summarized as follows.

\begin{itemize}
	\item We consider ``\textit{trading resources for privacy}". Specifically, to blur the real FoV, \textit{camouflaged} tile requests are added around the FoV. Then, the overall tile requests become a mixture of the tile requests overlapped with the FoV of a user and the camouflaged tile requests. We define spatial degree of privacy (SDoP) as a normalized number of camouflaged tile requests to reflect the privacy requirement. To satisfy a user with a larger SDoP requirement, more extra tiles are rendered and transmitted beyond those really requested, which makes inferring the real FoV more difficult at the cost of consuming more resources.
	\item To show the impact of privacy requirement, we optimize the durations for prediction, communication, and computing under arbitrarily given predictor, communication and computing resources to maximize the QoE. From the optimal solution, we find that longer total duration for communication and computing is required for a user with large value of SDoP, which degrades the performance for predicting tile requests. 
	Simulation with the state-of-the-art predictors on a real dataset shows that the increase of consumed resources for a user with a higher SDoP is larger than the degradation of prediction performance, hence the QoE can be improved.  
	
\end{itemize}


\section{System Model}\label{section-system_model}
\vspace{-.7mm}
Consider a tile-based VR video streaming system with a multi-access edge computing (MEC) server co-located with a base station (BS) that serves $K$ users.
The MEC server accesses a VR video library by local caching or high-speed backhaul, thus the delay from Internet to the MEC server can be neglected. The server also equips with powerful computing units for rendering, training a tile predictor and making the prediction.  
Each user requests VR videos from the library according to their own interests. Hence, the requests at the MEC server can be regarded as multiple individual requests.
In the sequel, we consider arbitrary one request for a video from a user.

Each VR video consists of $L$ segments in temporal domain, and each segment consists of $M$ tiles in spatial domain. The playback duration of each tile equals to the playback duration of a segment, denoted by $T_{\mathrm{seg}}$ \cite{optimizing_VR,survey_Hsu}. During the playback of a segment, the user turns around freely to view FoVs. 
Tiles overlapped with FoVs are the requested tiles. 

Each user equips with an head mounted display (HMD), which can measure the trace of viewpoint (centre point of the FoV) and log the tile requests,
upload the viewpoint trace or the tile requests to the MEC server, and pre-buffer segments. The HMD can also equipped with a light-weighted computing unit for training a predictor and making the prediction.

\subsection{Proactive VR Video Streaming Procedure}

Proactive VR video streaming requires to  predict the tile requests, either in direct or indirect manner. For direct prediction, the future tile requests are predicted from the observed tile requests \cite{Fixation_Prediction}. For indirect prediction, the future viewpoints are first predicted from the observed viewpoint trace \cite{optimizing_VR} and then mapped to the predicted tile requests \cite{TRACK,Xing_VR_Shannon}.



If a predictor needs to be trained in advance, the whole proactive procedure contains two stages: (1) offline predictor training, (2) online tile predicting, rendering and delivering. For some simple predictors with no need for training as those used in \cite{optimizing_VR,TRACK}, stage (1) can be omitted. 

Predictor training can be operated at the MEC server or the HMD. When training at the MEC server (i.e., centralized training), the real viewpoint traces or tile requests of multiple users should be uploaded. When training at each HMD (i.e., federated training), the real viewpoint traces or tile requests can be stored at each HMD and only the model parameters of predictor are uploaded to the MEC server.

\begin{figure}[htbp]
	\vspace{-0.4cm}
	\centering
	\begin{minipage}[t]{0.45\textwidth}
		\includegraphics[width=0.9\textwidth]{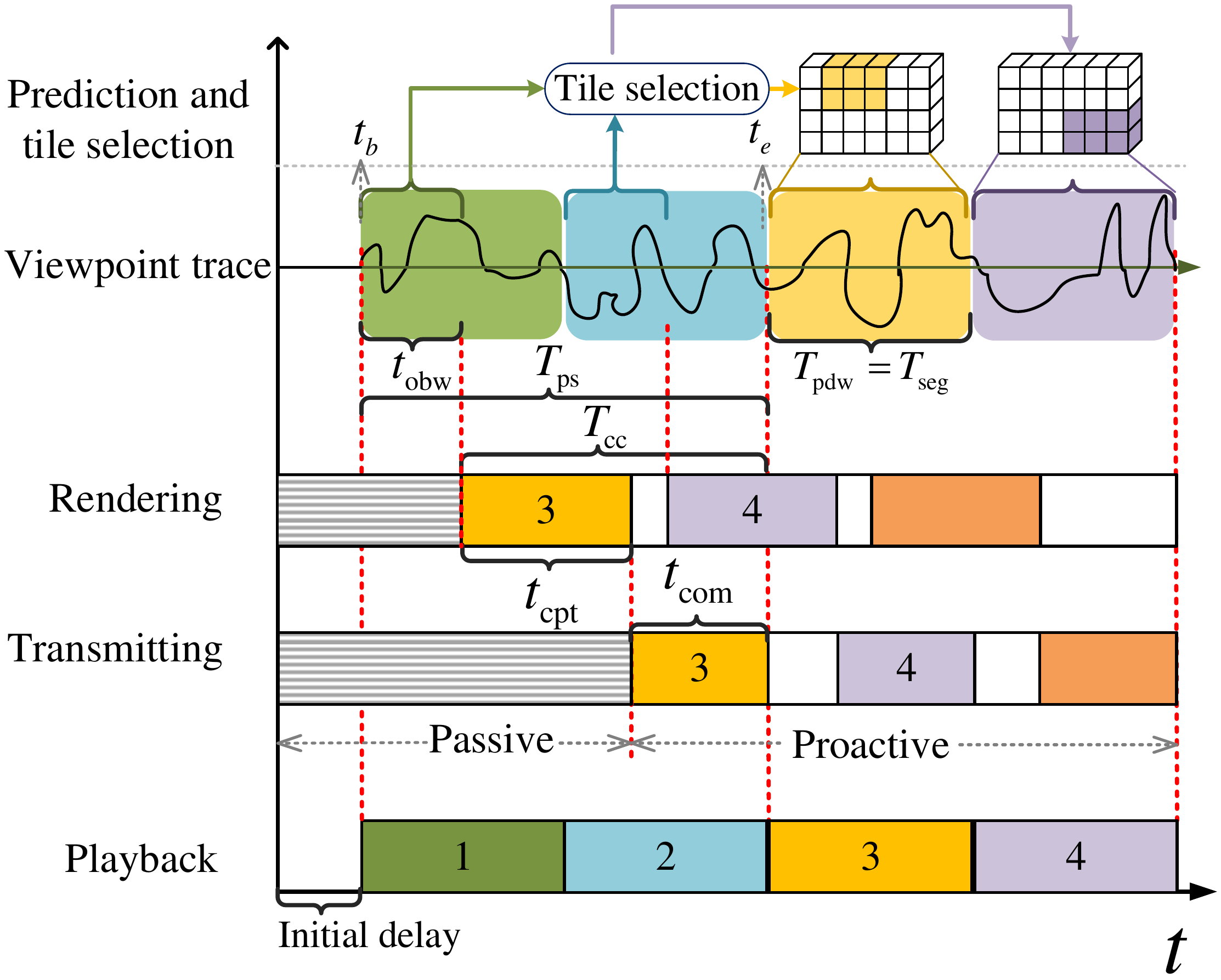}
	\end{minipage}
	\caption{Streaming the first four segments of a VR video, indirect prediction. $t_b$ is the start time of the observation window, $t_e$ is the start time of playback of the $l_0$th segment.}
	\label{Fig:MEC4VR_pipeline}
	\vspace{-0.3cm}
\end{figure}

The procedure in the online streaming stage is shown in Fig. \ref{Fig:MEC4VR_pipeline}. When a user requests a VR video, the MEC server first renders and transmits the initial ($l_{0}-1$)th segments in a passive streaming mode \cite{transmission_mode-standard-update}. After an initial delay, the first segment begins to play at the time instant $t_b$, which is the start time of an observation window. Then, proactive streaming begins, where subsequent segments are predicted with the well-trained predictor, and the predicted tiles are computed and transmitted. In the sequel, we take online proactive streaming for the $l_0$th segment as an example for elaboration.

After the viewpoint trace or the tile requests in the observation window with duration $t_{\mathrm{obw}}$ is collected, the tiles to be played in a prediction window with duration $T_{\mathrm{pdw}}=T_{\mathrm{seg}}$
can be predicted at the MEC server or a HMD.
If predicting viewpoints at the HMD, then the HMD can map the viewpoints to the tile requests and upload the tile requests, or it can leave the mapping to the MEC server and upload the viewpoints.

Based on the predicted tiles, 
the MEC server determines the tiles to be streamed, renders the tiles with duration $t_{\mathrm{cpt}}$, and transmits the tiles with duration $t_{\mathrm{com}}$. 
To avoid stalling, the tiles in the $l_0$th segment should be rendered and delivered before
the start time of playback of the $l_0$th segment, i.e., the time instant $t_e$. The duration starting from $t_b$ and terminating at $t_e$ is the online proactive streaming time for a segment, denoted as $T_{\mathrm{ps}}$. We can observe from the figure that $T_{\mathrm{ps}}=(l_0 -1)T_{\mathrm{seg}}$, where in the example $l_{0}=3$ and hence $T_{\mathrm{ps}}=2T_{\mathrm{seg}}$. The durations for observation, computing, and transmitting should satisfy $t_{\mathrm{obw}} + t_{\mathrm{cpt}} + t_{\mathrm{com}}=T_{\mathrm{ps}}$. The total duration for communication and computing is denoted as $t_{\mathrm{cc}}\triangleq t_{\mathrm{com}} + t_{\mathrm{cpt}}$.

\subsection{Computing and Transmission Model}
The number of bits that can be rendered per second, referred to as the computing rate, is $C_{\mathrm{cpt},k} \triangleq \frac{\mathcal{F}_{\mathrm{cpt}}}{K\cdot\mu_r}$ (in bit/s),
where $\mathcal{F}_{\mathrm{cpt}}$ is the configured resource at the MEC server for rendering  (in floating-point operations per second, FLOPS),
$\mu_r$ is the required floating-point operations (FLOPs) for rendering one bit of FoV (in FLOPs/bit) \cite{Xing_VR_Shannon}.

The BS serves $K$ single-antenna users using zero-forcing beamforming with $N_t$ antennas.
The instantaneous data rate at the $i$th time slot for the $k$th user is $C_{\mathrm{com},k}^{i}=B\log_2 \left(1+\frac{p_k^i d_k^{-\alpha}|\tilde{h}^i_k|^2}{\sigma^2} \right)$
where $B$ is the bandwidth, $\tilde{h}^i_k\triangleq (\mathbf{h}^i_k)^H\mathbf{w}^i_k$ is the equivalent channel gain, $p_k^i$ and $\mathbf{w}^i_k$ are respectively the transmit power and beamforming vector for the $k$th user,  $d_k$ and $\mathbf{h}^i_k\in\mathbb{C}^{N_t}$ are respectively the distance and the small scale channel vector from the BS to the $k$th user, $\alpha$ is the path-loss exponent, $\sigma^2$ is the noise power, and $(\cdot)^{H}$ denotes conjugate transpose.

We consider indoor users as in the literature, where the distances of users, $d_k$, usually change slightly \cite{survey_Hsu,NTHU_dataset} and hence are assumed fixed.
Due to the head movement and the variation of the environment,
small-scale channels are time-varying, which are assumed as remaining constant in each time slot with duration $\Delta T$ and changing independently with identical distribution among time slots. With the proactive transmission, the rendered tiles in a segment should be transmitted with duration $t_{\mathrm{com}}$. The number of bits transmitted with $t_{\mathrm{com}}$ can be expressed as $\overline{C}_{\mathrm{com},k}\cdot t_{\mathrm{com}}$, where  $\overline{C}_{\mathrm{com},k} \triangleq \frac{1}{N_s}\sum_{i=1}^{N_s}C_{\mathrm{com},k}^{i} \cdot \Delta T$
is the time average transmission rate, and $N_s$ is the number of time slots in $t_{\mathrm{com}}$.
Since future channels are unknown at the time instant $t_b$, we use ensemble-average rate $\mathbbm{E}_h\{C_{\mathrm{com},k}\}$ \cite{ergodic-capacity} to approximate the time-average rate $\overline{C}_{\mathrm{com},k}$, where $\mathbbm{E}_h\{\cdot\}$ is the expectation over $h$, which is accurate when $N_s$ or $N_t/K$ is large \cite{Xing_VR_Shannon}. By allocating transmit power among users to compensate the path loss, the ensemble-average transmission rate for each user is equal.
For notional simplicity, we use $C_{\mathrm{com}}$ and $C_{\mathrm{cpt}}$ to replace $\mathbbm{E}_h\{C_{\mathrm{com},k}\}$ and $C_{\mathrm{cpt},k}$. 


We  use the ratio of tiles in a segment that can be rendered and transmitted with assigned transmission and computing rates and corresponding durations to measure the communication and computing (CC) capability of the system for streaming the tiles, which is,
\begin{align}\label{def:C_cc}
	C_{\mathrm{cc}}(t_{\mathrm{com}}, t_{\mathrm{cpt}}) \triangleq \left. \min\left\{\frac{C_{\mathrm{com}}t_{\mathrm{com}}}{s_{\mathrm{com}}^{}}, \frac{C_{\mathrm{cpt}}t_{\mathrm{cpt}}}{s_{\mathrm{cpt}}^{}}, M\right\} \middle/ M \right.
\end{align}
where $\min\left\{\frac{C_{\mathrm{com}}t_{\mathrm{com}}}{s_{\mathrm{com}}^{}}, \frac{C_{\mathrm{cpt}}t_{\mathrm{cpt}}}{s_{\mathrm{cpt}}^{}}, M\right\}$ is the number of tiles that can be computed and transmitted, $s_{\mathrm{com}} = {px}_w \cdot{px}_h \cdot b \cdot r_f \cdot T_{\mathrm{seg}}/\gamma_{c}$ \cite{HuaWei_Cloud_VR} is the number of bits in each tile for transmission, $s_{\mathrm{cpt}} = {px}_w \cdot{px}_h \cdot b \cdot r_f \cdot T_{\mathrm{seg}}$ is the number of bits in a tile for rendering, ${px}_w$ and ${px}_h$ are the pixels in wide and high of a tile, $b$ is the number of bits per pixel relevant to color depth \cite{HuaWei_Cloud_VR}, $r_f$ is the frame rate, and $\gamma_c$ is the compression ratio.

To reflect the system capability of streaming tiles in unit time, we further define the \textit{resources rate} as
\begin{align}
	R_{\mathrm{cc}}\triangleq \frac{C_{\mathrm{cc}}(t_{\mathrm{com}}, t_{\mathrm{cpt}})}{t_{\mathrm{cc}}}\label{def:R_cc}
\end{align}

%
%
%
%

%

\subsection{Performance Metric of Tile Prediction}\label{section:DoO_def}

Average segment degree of overlap (average-DoO) has been used to measure the prediction performance for a VR video \cite{Xing_VR_Shannon}. It indicates the average overlap of the predicted tiles and the real requested tiles among all the proactively streamed segments, which is defined as
\begin{align}\label{def:DoO}
	\mathcal{D}(t_{\mathrm{obw}}) \triangleq \frac{1}{L-l_0 + 1}\sum_{l=l_0}^L\frac{\mathbf{q}_{l}^\mathsf{T}\cdot\mathbf{e}_{l}({t_{\mathrm{obw}}})  }{\|\mathbf{q}_{l}\|_1}\in[0,100\%]
\end{align}
where $\mathbf{q}_{l}\triangleq [q_{l,1},...,q_{l,M}]^\mathsf{T}$ denotes the real tile requests for the $l$th segment with $q_{l,m}\in\{0,1\}$, $\mathbf{e}_{l}({t_{\mathrm{obw}}})\triangleq [e_{l,1}({t_{\mathrm{obw}}}),...,e_{l,M}({t_{\mathrm{obw}}})]^\mathsf{T}$ denotes the predicted tile requests for the segment with $e_{l,m}({t_{\mathrm{obw}}})\in\{0,1\}$, and
$(\cdot)^\mathsf{T}$ denotes transpose. When the $m$th tile in the $l$th segment is truly requested, $q_{l,m}=1$, otherwise $q_{l,m}=0$. When the tile is predicted to be requested, $e_{l,m}({t_{\mathrm{obw}}})=1$, otherwise it is zero. We consider $\|\mathbf{e}_{l}\left(t_{\mathrm{obw}}\right)\|_1 = N_{\textit{fov}}$, where $N_{\textit{fov}}$ is the number of tiles in a FoV  and $\|\cdot\|_1$ denotes the $\ell_1$ norm of a vector.

A larger value of average-DoO indicates a better prediction.


As verified in \cite{Xing_VR_Shannon}, a predictor can be more accurate with a longer observation window. Therefore, average-DoO is a monotonically increasing function of $t_{\mathrm{obw}}$. 

\subsection{Metric of Quality of Experience}
For proactive tile-based VR video streaming, the QoE can be measured by the percentage of the correctly delivered tiles among all the really requested tiles \cite{Xing_VR_Shannon}, i.e.,
\begin{align}
	\mathrm{QoE} &= \frac{1}{L-l_0 + 1}\sum_{l=l_0}^L\frac{(\mathbf{q}_{l})^{\mathsf{T}}\cdot \mathbf{s}_{l}}{\|\mathbf{q}_{l}\|_1}\label{qoe_cal}
\end{align}
where $\mathbf{s}_l\triangleq [s_{l,1},...,s_{l,M}]^\mathsf{T}$ with  $s_{l,m}\in\{0,1\}$. When the $m$th tile in the $l$th segment is determined by the MEC server to be streamed, $s_{l,m}=1$, otherwise $s_{l,m}=0$.

From \eqref{def:C_cc} and \eqref{def:DoO}, the QoE  can be expressed as
\begin{align*}
	\mathrm{QoE} = \mathcal{Q}\left(\mathcal{D}\left(t_{\mathrm{obw}}\right), C_{\mathrm{cc}}(t_{\mathrm{com}}, t_{\mathrm{cpt}})\right)\in[0,100\%]
\end{align*}
which depends on the average-DoO and the CC capability.
When the value of the QoE is $100\%$, all the really requested tiles are proactively computed and delivered before playback. 
When $C_{\mathrm{cc}}(t_{\mathrm{com}}, t_{\mathrm{cpt}})$ is larger, more tiles can be rendered and transmitted, then more really requested tiles can be delivered. When $\mathcal{D}\left(t_{\mathrm{obw}}\right)$ is improved, more streamed tiles are the really requested tiles. This indicates that the QoE monotonically increases with the average-DoO and CC capability, respectively.

\vspace{-2mm}
\section{FoV Privacy-protecting Proactive Streaming}

\begin{table*}[htbp]
	\vspace{-0.7cm}
	\caption{FoV leakage in proactive VR video streaming procedure}\label{table:FoV_leakage_two_manners}
	\vspace{-0.3cm}
	\begin{center}
		\begin{tabular}{|p{1mm}<{\centering}|c|c|c|c|c|c|}
			\hline
			\multirow{2}{*}{\!\!No.\!\!\!\!\!}&\!\!\!Direct/Indirect\!\!\!&\multirow{2}{*}{Where to train and predict}&\multicolumn{2}{c|}{Predictor training} & \multicolumn{2}{c|}{Tile prediction}\\
			\cline{4-7}
			&prediction&& \!\!HMD upload data\!\!&\!\!\textbf{FoV leakage}\!\!&\!\!HMD upload data\!\!&\!\!\textbf{FoV leakage}\!\!\\
			\hline
			1&\multirow{6}{*}{Direct}&Training at MEC, predicting at MEC&Real tile requests& \checkmark & Real tile requests & \checkmark\\
			\cline{1-1}\cline{3-7}
			2&&Training at MEC, predicting at HMD&Real tile requests& \checkmark & Predicted tile requests & \ \checkmark$^*$\\
			\cline{1-1}\cline{3-7}
			3&&Training at HMD, predicting at HMD&Model parameters & $\times$ & Predicted tile requests & \ \checkmark$^*$\\
			\cline{1-1}\cline{3-7}
			4&&Training at HMD, predicting at MEC&Model parameters & $\times$ & Real tile requests & \checkmark\\
			\cline{1-1}\cline{3-7}
			5&&No need for training, predicting at MEC& \backslashbox[1.7cm]{}{} & \backslashbox[1.7cm]{}{} & Real tile requests &\checkmark\\
			\cline{1-1}\cline{3-7}
			6&&No need for training, predicting at HMD& \backslashbox[1.7cm]{}{} & \backslashbox[1.7cm]{}{}& Predicted tile requests & \ \checkmark$^*$\\
			\hline
			7&\multirow{6}{*}{Indirect}&Training at MEC, predicting at MEC&Real viewpoints& \checkmark & Real viewpoints & \checkmark\\
			\cline{1-1}\cline{3-7}
			8&&Training at MEC, predicting at HMD&Real viewpoints& \checkmark & Predicted viewpoints or tile requests & \ \checkmark$^*$\\
			\cline{1-1}\cline{3-7}
			9&&Training at HMD, predicting at HMD&Model parameters & $\times$ & Predicted viewpoints or tile requests & \ \checkmark$^*$\\
			\cline{1-1}\cline{3-7}
			10 \ &&Training at HMD, predicting at MEC&Model parameters & $\times$ & Real viewpoints& \checkmark\\
			\cline{1-1}\cline{3-7}
			11&&No need for training, predicting at MEC& \backslashbox[1.7cm]{}{} & \backslashbox[1.7cm]{}{} & Real viewpoints &\checkmark\\
			\cline{1-1}\cline{3-7}
			12&&No need for training, predicting at HMD& \backslashbox[1.7cm]{}{} & \backslashbox[1.7cm]{}{}& Predicted viewpoints or tile requests & \ \checkmark$^*$\\
			\hline
		\end{tabular}
	\end{center}
	\footnotesize{ \ \ \ $^*$When prediction is accurate.}
	\vspace{-0.1cm}
\end{table*}

In the proactive VR Video streaming procedure, the FoV may be leaked out during predictor training and online prediction. According to whether the tile requests are predicted directly or indirectly, whether a predictor requires training, and where the training and predicting are respectively executed, we provide  twelve cases in Table \ref{table:FoV_leakage_two_manners}.

We can find that although federated training or simple predictors with no need for training can avoid
the FoV leakage during the predictor training stage, the real FoV can still be inferred at the MEC server during the online prediction, because either the real or predicted effective data for inferring (viewpoints and tile requests) is uploaded. 
In this section, we provide a method to avoid the FoV leakage, define a metric to reflect the privacy requirement for FoV, and discuss how to satisfy the requirement.

\subsection{Conceal Real FoVs with Extra Tiles}

The FoV leakage comes from the uploaded viewpoints or tile requests.
To protect the FoVs, the HMD should conceal the real data. If this is achieved by adding noise as in  \cite{privacy-preserving_eye_tracking_2021,privacy_def_eye_track}, then the tiles recovered from the noisy viewpoints cannot include all the really-requested tiles and black holes occur. Alternatively, we can
remove some real tile requests or add some camouflaged tile requests. If some really-requested tiles are not rendered and transmitted, again, black holes occur. Therefore, we add camouflaged tile requests. The overall tiles to be rendered and delivered are composed of the predicted tile requests from a user (with the number as $N_{\textit{fov}}$) and the camouflaged tile requests (with the number denoted as $N_{\textit{cf}}$).

\begin{figure}[htbp]
	\vspace{-0.2cm}
	\centering
	\subfloat[No privacy requirement ($\rho_{s}=0$), $N_{\textit{fov}}=4\times 2=8$.]{\label{Fig:sDoP_explain_0}
		\begin{minipage}[c]{1\linewidth}
			\centering
			\includegraphics[width=1\textwidth]{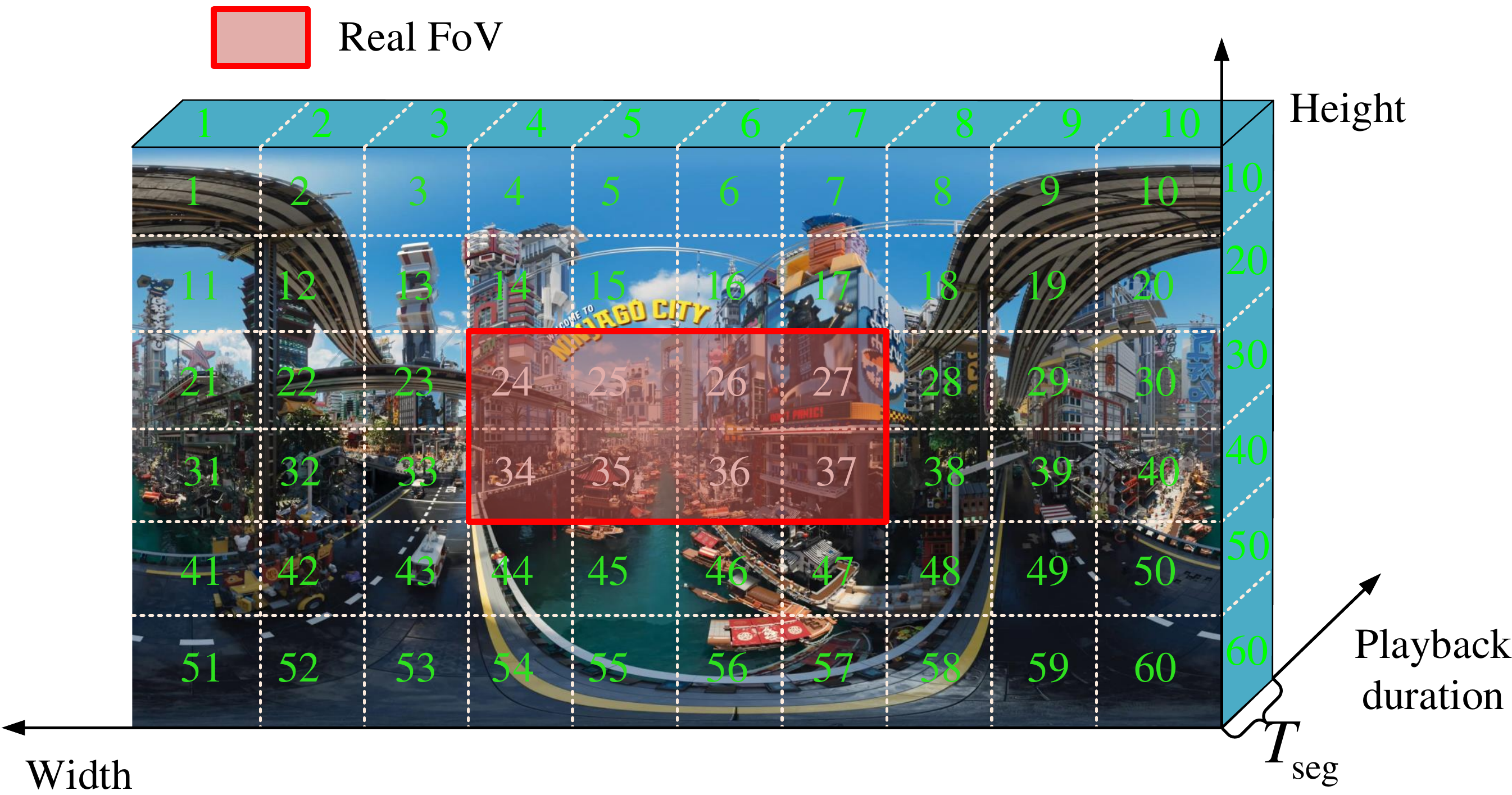}
		\end{minipage}
	}\vspace{-0.2cm}\\
	\subfloat[Low privacy requirement ($\rho_{s}=31\%$), $N_{\textit{cf}}=6\times 4 - N_{\textit{fov}} =16$.]{\label{Fig:sDoP_explain_1}
		\begin{minipage}[c]{1\linewidth}
			\centering
			\includegraphics[width=1\textwidth]{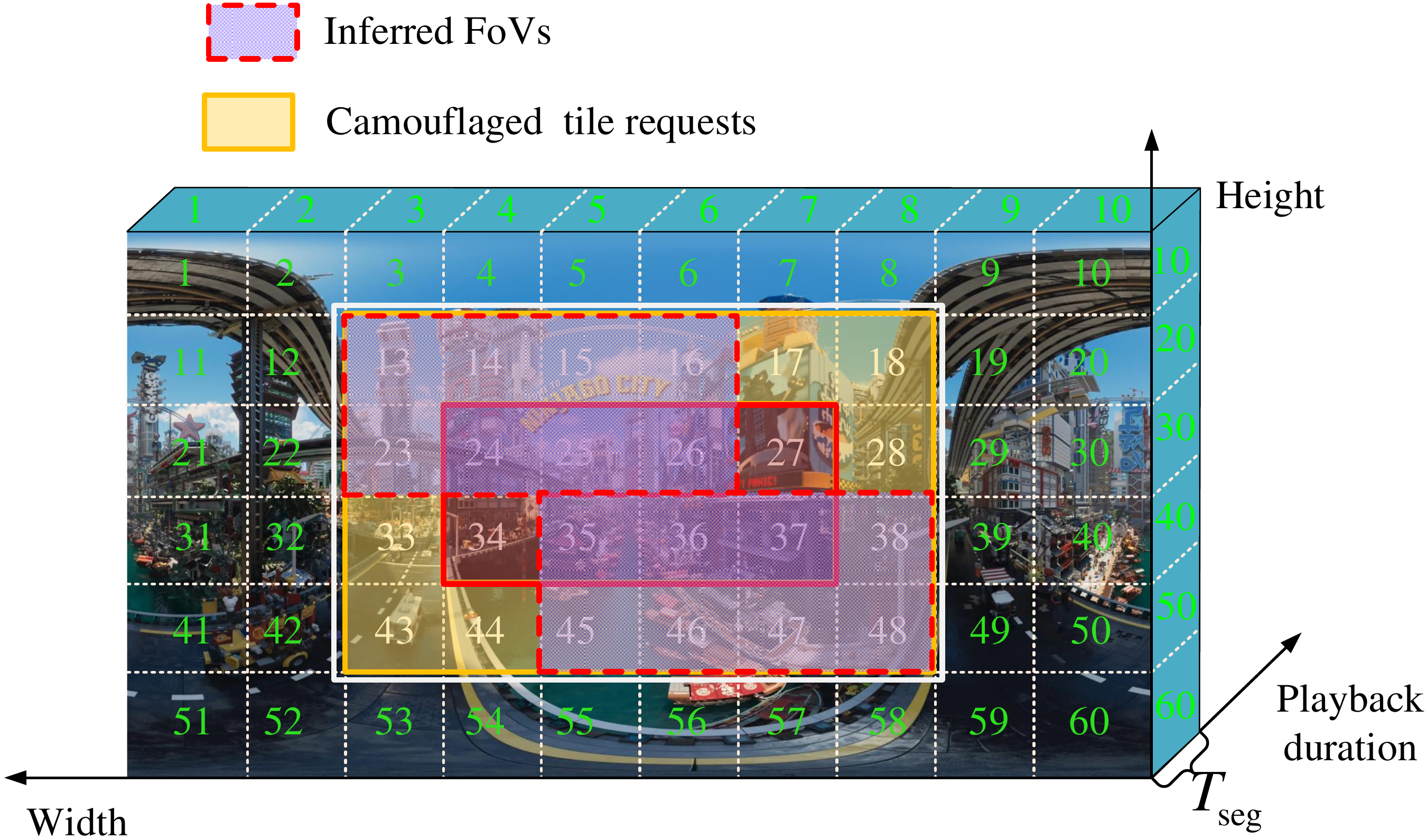}
		\end{minipage}
	}\vspace{-0.4cm}\\
	\subfloat[High privacy requirement ($\rho_{s}=65\%$), $N_{\textit{cf}}=7\times 6 - 8 =34$.]{\label{Fig:sDoP_explain_2}
		\begin{minipage}[c]{1\linewidth}
			\centering
			\includegraphics[width=1\textwidth]{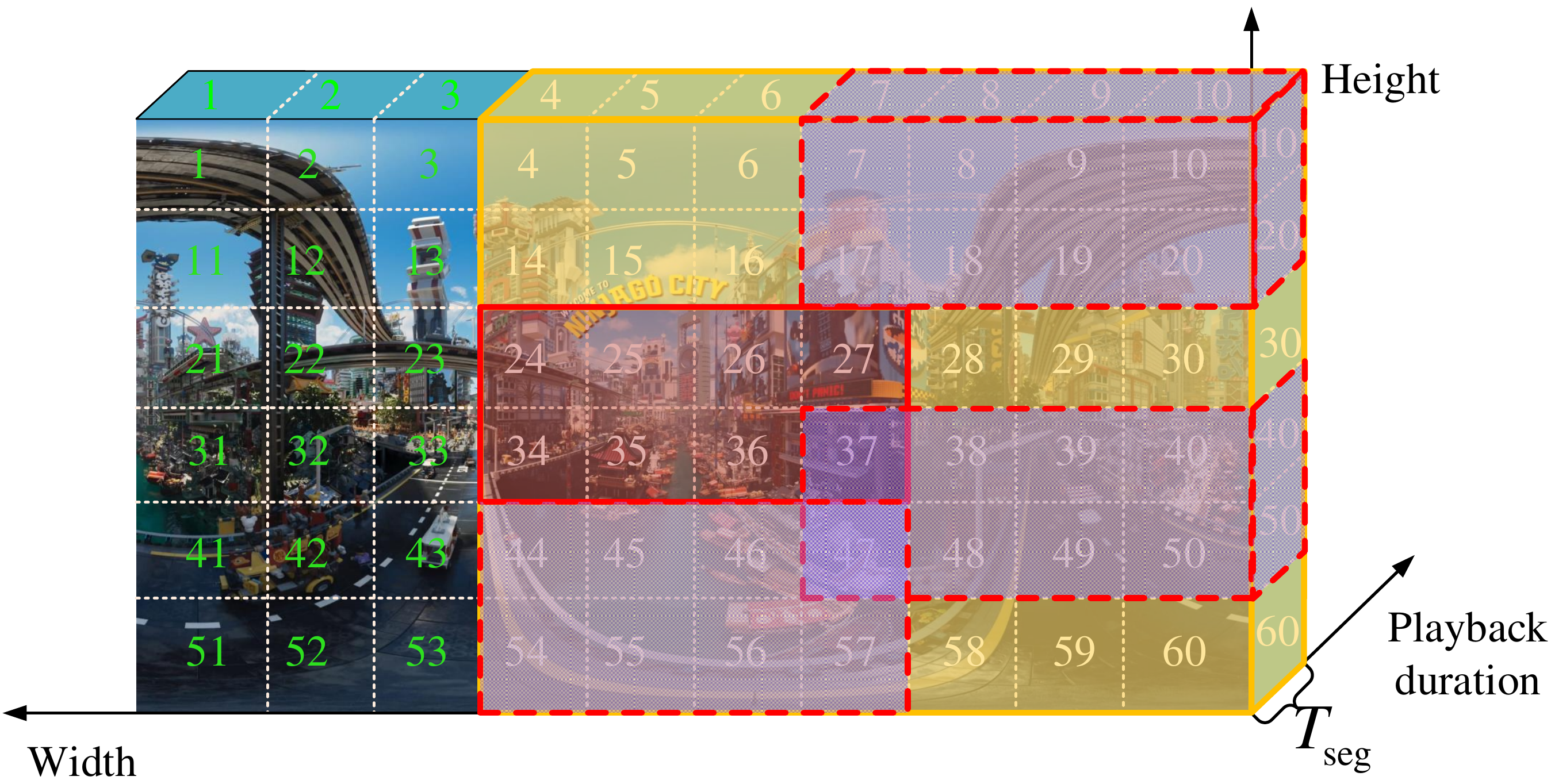}
		\end{minipage}
	}\vspace{-0.1cm}
	\caption{Real FoV, camouflaged tile requests, and inferred FoVs with different $\rho_{s}$ in a segment, $M=60$.}\label{Fig:sDoP_explain}
	\vspace{-0.4cm}
\end{figure}

As shown in Fig. \ref{Fig:sDoP_explain_0}, without the camouflaged tile requests, when the MEC server receives the real tile requests with No. 24-27, 34-37, it can immediately obtain the content in the real FoV. As shown in Fig. \ref{Fig:sDoP_explain_1}, if the camouflaged tile requests with No. 13-18, 23, 28, 33, 38, 43-48 are added, the overall tiles to be streamed are with No. 13-18, 23-28, 33-38, 43-48 and the MEC server only sees the white borderline region.
As the number of camouflaged tile requests increases, it becomes more and more difficult for the server to correctly infer the real FoV. In Fig. \ref{Fig:sDoP_explain_2}, some inferred FoVs might even have no intersection with the real FoV.

With camouflaged tile requests, we summarize the FoV protection in the twelve cases in Table \ref{table:FoV_protecting_two_manners}.  In most of cases, the FoV can be protected, except several indirect prediction cases where the real viewpoints are uploaded, which should not be used in practice.

\begin{table*}[htbp]
	\vspace{-0.15cm}
	\caption{FoV protection with camouflaged tile requests}\label{table:FoV_protecting_two_manners}
	\vspace{-0.3cm}
	\begin{center}
		\begin{tabular}{|c|c|c|c|c|}
			\hline
			\multirow{2}{*}{\!\!No.\!\!\!\!\!}&\multicolumn{2}{c|}{Predictor training} & \multicolumn{2}{c|}{Tile prediction}\\
			\cline{2-5}
			& \!\!HMD upload data\!\!&\!\!\textbf{FoV protection}\!\!&\!\!HMD upload data\!\!&\!\!\textbf{FoV protection}\!\!\\
			\hline
			1&Real and camouflaged tile requests& \multirow{6}{*}{\Large{\checkmark}} & Real and camouflaged tile requests & \multirow{6}{*}{\Large{\checkmark}}\\
			\cline{1-2}\cline{4-4}
			2&Real and camouflaged tile requests&  & Predicted and camouflaged tile requests & \\
			\cline{1-2}\cline{4-4}
			3&Model parameters &  & Predicted and camouflaged tile requests & \\
			\cline{1-2}\cline{4-4}
			4&Model parameters &  & Real and camouflaged tile requests & \\
			\cline{1-2}\cline{4-4}
			5& \backslashbox[1.7cm]{}{}&  & Real and camouflaged tile requests &\\
			\cline{1-2}\cline{4-4}
			6& \backslashbox[1.7cm]{}{}& & Predicted and camouflaged tile requests & \\
			\hline
			7&Real viewpoints& $\times$ & Real viewpoints & $\times$\\
			\hline
			8&Real viewpoints& $\times$ & Predicted and camouflaged tile requests$^\#$ & \checkmark\\
			\hline
			\bf{9}&Model parameters & \checkmark & Predicted and camouflaged tile requests$^\#$ & \bf{\checkmark}\\
			\hline
			10 &Model parameters & \checkmark & Real viewpoints& $\times$\\
			\hline
			11& \backslashbox[1.7cm]{}{}& \checkmark & Real viewpoints &$\times$\\
			\hline
			\bf{12}& \backslashbox[1.7cm]{}{} & \checkmark& Predicted and camouflaged tile requests$^\#$ &  \checkmark\\
			\hline
		\end{tabular}
	\end{center}
	\footnotesize{ \ \ \ \ \ \ \ \ \ \ \ \ \ \ \ \ \ \ $^\#$The HMD uploads the predicted tile requests rather than viewpoints to protect FoVs.}
	\vspace{-0.45cm}
\end{table*}

\subsection{Spatial Degree of Privacy}

To reflect the privacy requirement of a user for protecting the real FoV during proactive VR streaming procedure, we define a metric of \textit{spatial degree of privacy}, which is the normalized number of camouflaged tile requests, i.e.,
\begin{equation}\label{def:sDoP}
	\rho_{s} \triangleq \frac{N_{\textit{cf}}}{M - N_{\textit{fov}}}\in[0,100\%]
\end{equation}
When $\rho_s = 0$, the user has no privacy requirement, $N_{\textit{cf}}=0$, i.e., only the real or predicted tile requests are uploaded to the MEC server.
When $\rho_s = 100\%$, the user has most stringent privacy requirement, $N_{\textit{cf}} + N_{\textit{fov}}=M$, 
i.e., the MEC server sees the whole panoramic region and cannot infer the real FoV. When a user demands for a VR video, $\rho_{s}$ is determined by the user. Given the required SDoP, the overall number of tiles in a segment to be rendered and transmitted is
\begin{align}\label{def:N_p0}
	N_p(\rho_s) =  N_{\textit{fov}} + N_{\textit{cf}} =  N_{\textit{fov}} + \lceil\rho_s (M - N_{\textit{fov}})\rceil
\end{align}
where $\lceil\cdot\rceil$ is the ceiling function.

\vspace{-0.5mm}
\subsection{Satisfy the SDoP Requirement}
\vspace{-0.5mm}
If the CC capability is not sufficient for rendering and transmitting the overall tiles  of a user with required SDoP, i.e., $ C_{\mathrm{cc}}(t_{\mathrm{com}} ,t_{\mathrm{cpt}})\cdot M < N_p(\rho_s)$, then the MEC server will still stream the predicted tiles to save resources, i.e., $N_{p}(\rho_s) = N_{\textit{fov}}$. As a consequence, the FoV will be leaked out. This suggests that the CC capability should be larger for a user with larger value of $\rho_{s}$. To gain useful insight, we assume that the CC capability can exactly satisfy the required SDoP, i.e., $C_{\mathrm{cc}}(t_{\mathrm{com}} ,t_{\mathrm{cpt}})\cdot M = N_p(\rho_s)$. 

\vspace{-0.05cm}
\section{Impacts of SDoP on the QoE}
\vspace{-0.1cm}
In this section, we investigate the impact of SDoP requirement on the QoE. To this end, we first analyze the impact of SDoP on the duration for communication and computing and average-DoO, respectively, from the jointly optimized durations of observation window, communication, and computing to maximize the QoE. Then, we discuss the overall impact of SDoP on the QoE.

\vspace{-1mm}
\subsection{Joint Optimization of the Durations}
\vspace{-1mm}
Given computing rate $C_{\mathrm{cpt}}$, transmission rate $C_{\mathrm{com}}$ and arbitrary  SDoP $\rho_s$, we can find the optimal durations for observation window, communication and computing to maximize the QoE from the following problem
\begin{subequations}
	\begin{align}
		\textbf{P0}:&\ \ \ \  \ \max_{t_{\mathrm{obw}}, t_{\mathrm{cpt}},t_{\mathrm{com}}}  \mathcal{Q}\left(\mathcal{D}\left(t_{\mathrm{obw}}\right), C_{\mathrm{cc}}(t_{\mathrm{com}}, t_{\mathrm{cpt}})\right) \\
		&  \ \ \ \  \  s.t.  \ \   C_{\mathrm{cc}}(t_{\mathrm{com}}, t_{\mathrm{cpt}}) = \frac{N_{p}(\rho_s)}{M} \label{P0_C_cc}\\
		& \ \ \ \ \  \  \ \  \ \ \ t_{\mathrm{obw}} + t_{\mathrm{cpt}} + t_{\mathrm{com}} = T_{\mathrm{ps}} \label{P0_t_ps}
	\end{align}
\end{subequations}
As derived in the appendix of \cite{Xing_spatial_DoP_VR_streaming}, the solution of \textbf{P0} is,
\begin{subequations}\label{P3:opt_solution_general}
	\begin{align}
		&t_{\mathrm{obw}}^{*}= \left\lfloor\left.  \left(T_{\mathrm{ps}} - \left(\frac{s_{\mathrm{com}}}{C_{\mathrm{com}}} + \frac{s_{\mathrm{cpt}}}{C_{\mathrm{cpt}}}\right)\cdot N_p(\rho_s)\right)\middle/\tau\right.\right\rfloor
		\label{P3:opt_solution_general_t_obw}\\
		&t_{\mathrm{com}}^{*}=\frac{s_{\mathrm{com}}N_p(\rho_s)}{C_{\mathrm{com}}} \ \ t_{\mathrm{cpt}}^{*}=\frac{s_{\mathrm{cpt}}N_p(\rho_s)}{C_{\mathrm{cpt}}}\label{P3:opt_solution_general_t_cc}
	\end{align}
\end{subequations}
where $\tau$ (in seconds) is the sampling interval in the observation window. The optimal duration for communication and computing $t_{\mathrm{cc}}^*$ can be obtained from \eqref{P3:opt_solution_general_t_cc} as
\begin{align}
	t_{\mathrm{cc}}^* = \left(\frac{s_{\mathrm{com}}}{C_{\mathrm{com}}} + \frac{s_{\mathrm{cpt}}}{C_{\mathrm{cpt}}}\right)\cdot N_p(\rho_s)\label{opt_t_cc}
\end{align}
By substituting \eqref{P3:opt_solution_general_t_cc} into \eqref{def:R_cc}, the maximized  resources rate is
\begin{align}
	R_{\mathrm{cc}}^* = \left. 1 \middle/ \left(\frac{s_{\mathrm{com}}M}{C_{\mathrm{com}}} + \frac{s_{\mathrm{cpt}}M}{C_{\mathrm{cpt}}}\right) \right.\label{opt_R_cc}
\end{align}
where $\left(\frac{s_{\mathrm{com}}M}{C_{\mathrm{com}}} + \frac{s_{\mathrm{cpt}}M}{C_{\mathrm{cpt}}}\right)$ is the  required optimal duration to render and transmit all tiles in a segment.

\subsection{Impact of SDoP}
By substituting \eqref{opt_t_cc} and \eqref{opt_R_cc} into \eqref{def:R_cc}, the CC capability can be rewritten as $C_{\mathrm{cc}}(t_{\mathrm{com}}^*, t_{\mathrm{cpt}}^*) = R_{\mathrm{cc}}^* t_{\mathrm{cc}}^*$. Further considering \eqref{opt_t_cc} and \eqref{opt_R_cc}, we can find that with a large value of $\rho_s$, CC capability can be increased by using longer total duration for communication and computing $t_{\mathrm{cc}}^*$. Then, the duration of observation window $t_{\mathrm{obw}}^* = T_{\mathrm{ps}} - t_{\mathrm{cc}}^*$ will be reduced, which can also be verified from \eqref{P3:opt_solution_general_t_obw}. The reduction of $t_{\mathrm{obw}}^*$ degrades the average-DoO.
A large value of SDoP needs a higher CC capability, which however degrades the average-DoO. The overall impact on the QoE depends on if the QoE is dominated by the increase of CC capability or the reduction of the average-DoO.

\vspace{-0.1cm}
\section{Trace-Driven Simulation Results}

\begin{table*}[htbp]
	\captionsetup{font={small}}
	\vspace{-0.5cm}
	\caption{Settings of VR videos}\label{table:LR}
	\vspace{-0.4cm}
	\begin{center}
		\begin{tabular}{|c|c|c|c|}
			\hline
			Resolution & 3840$\times$2160 pixels\cite{FoV_aware_tile} &Bits per pixel &$12$ \cite{HuaWei_Cloud_VR}\\
			\hline
			Number of tiles $M$ & 10 rows $\times$ 20 columns = 200\cite{NTHU_dataset} & Frame rate $r_f$ & 30 FPS \cite{FoV_aware_tile} \\ \hline
			Pixels in wide of a tile ${px}_w$ & $3840/20 = 192$&Pixels in height of a tile ${px}_h$ & $2160/10=216$\\ \hline
			Compression ratio $\gamma_c$ & 2.41\cite{HEVC_lossless_coding}& Playback duration of a segment $T_{\mathrm{seg}}$ & 1 s \cite{FoV_aware_tile}\\ \hline
			Number of bits in a tile for transmission $s_{\mathrm{com}}$ & 5.9 Mbits & Number of bits in a tile for rendering $s_{\mathrm{cpt}}$ & 14.2 Mbits\\ \hline
			Size of FoV & $100^{\circ}\times100^{\circ}$ circles \cite{FoV_size_NTHU,TRACK} & Number of tiles in a FoV $N_{\textit{fov}}$ & 33\cite{NTHU_dataset}\\ \hline
		\end{tabular}
	\end{center}
	\vspace{-0.4cm}
\end{table*}

In this section, we show the overall impact of SDoP on QoE via trace-driven simulation results.

First, we consider the tile prediction on a real dataset \cite{NTHU_dataset}, where 300 traces of head movement positions from 30 users watching 10 VR videos are used for training and testing predictors.
\footnote{According to the analysis in  \cite{TRACK,Romero_ACM_MMsys_20_paper}, the traces of the first 20 users in the dataset have mistakes, thus we only use the traces of the other 30 users.}
We randomly split the total traces into training and testing sets with ratio 8:2.

We use two indirect tile predictors, \textit{deep-position-only} and \textit{trivial-motion} predictors, which achieve the state-of-the-art accuracy for the dataset, according to evaluation in \cite{TRACK}. These predictors represent two types of predictors: The former needs to be trained while the latter is with no need for training.
The \textit{deep-position-only} predictor employs a sequence-to-sequence LSTM-based architecture, which uses the time series of past viewpoint positions as input to predict the time series of future positions \cite{TRACK}. The predictor does not consider the time required for computing and communication as well as the SDoP. To reserve time for computing and communication, we tailor the predictor by setting the duration between the end of the observation window and the beginning of the prediction windows as $t_{\mathrm{cc}}^*$, and setting the durations of observation and prediction windows as $t_{\mathrm{obw}}^*$ and $T_{\mathrm{seg}}$, respectively. To protect the FoV, we consider training and predicting at the HMD, which is case 9 in Table \ref{table:FoV_protecting_two_manners}. In predictor training, we consider a classical federated learning algorithm, \texttt{FederatedAveraging} \cite{google_federated_learning_17}. The settings of the federated learning are as follows. In each round, we select \textit{all} of $K=30$ users to update the model parameters of the predictor. The number of local epochs for each user is $E_l=50$, the number of communication rounds is $R=10$. Hence, every trace in the training set is used $50\times10=500$ times, which is consistent with the centralized training \cite{TRACK}. The weighting coefficient of the $k$th user on the model parameter is $c_k =\frac{n_k}{N_{\textit{train}}}$, where $N_{\textit{train}}=300\times0.8=240$ is the total number of traces in the training set, $n_k$ is the number of video traces of the $k$th user in the training set. Due to the random division of training and test sets, $n_k$ varies from 6 to 10.
We refer to the predictor as \textit{tailored federated position-only} predictor. Other details and hyper-parameters of the tailored predictor are the same as the \textit{deep-position-only} predictor \cite{TRACK}. The \textit{trivial-motion} predictor simply uses the last position in the observation window as the predicted time series of future positions \cite{TRACK}. We consider predicting at the HMD, which is case 12 in Table \ref{table:FoV_protecting_two_manners}.

The maximized resources rate $R_{\mathrm{cc}}^*$ depends on the configured communication and computing resources as well as the number of users. For example, when $K=4$, $N_t=8$, $P=24$ dBm, $B=150$ MHz, and $d_k=5$ m, the ensemble-average transmission rate for a user is $C_{\mathrm{com}}=2.85$ Gbps\cite{Xing_VR_Shannon}. When Nvidia RTX 8000 GPU is used for rendering VR videos for four users, the computing rate for a user is $C_{\mathrm{cpt}}=2.2$ Gbps \cite{Xing_VR_Shannon}. Then, the maximized resources rate is $R_{\mathrm{cc}}^* = \left. 1 \middle/ \left(\frac{s_{\mathrm{com}}M}{C_{\mathrm{com}}} + \frac{s_{\mathrm{cpt}}M}{C_{\mathrm{cpt}}}\right) \right. = 0.6$. To reflect the variation of configured resources, we set $R_{\mathrm{cc}}^*\in[0.6,2]$.

The settings of VR videos are listed in Table \ref{table:LR}.
To gain useful insight, we assume that all users have identical SDoP requirement for all videos, ranging from 0 to 100\%. After training the tailored federated position-only predictor, $\mathcal{D}(t_{\mathrm{obw}}^*)$ of the two predictors and $C_{\mathrm{cc}}(t_{\mathrm{com}}^*, t_{\mathrm{cpt}}^*)$ can be obtained. The QoE of the two predictors is first calculated from \eqref{qoe_cal}, and then averaged over the test set. The details of simulation procedure can be found in \emph{Procedure 1} of \cite{Xing_spatial_DoP_VR_streaming}.

\begin{figure}[htbp]
	\vspace{-0.1cm}
	\centering
	\begin{minipage}[t]{0.85\linewidth}
		\includegraphics[width=1\textwidth]{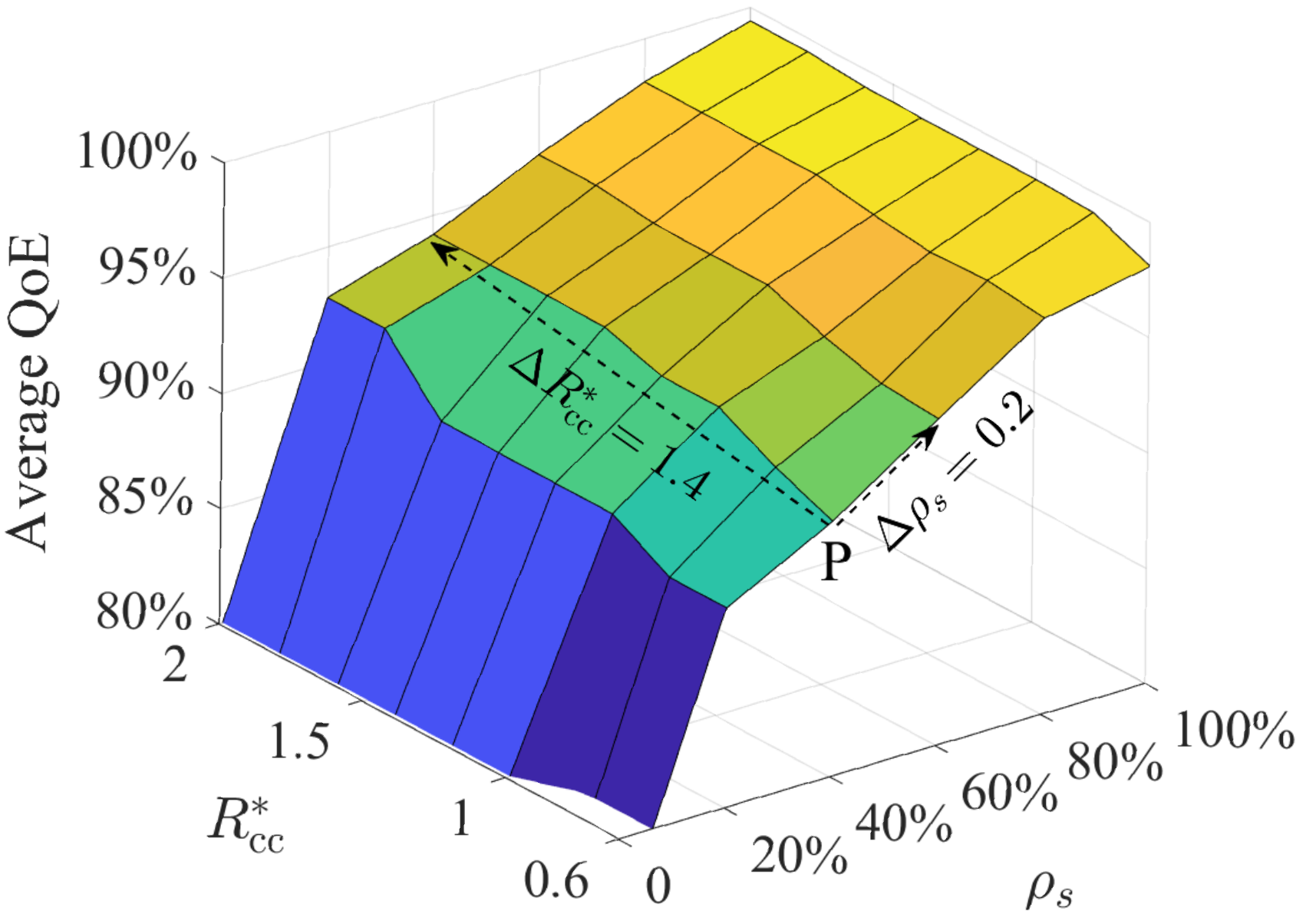}
	\end{minipage}
	\vspace{-0.2cm}
	\caption{Average QoE v.s. resource rates and SDoP.}\label{Fig:qoe_rho_T_cc_m}
\end{figure}


In Fig. \ref{Fig:qoe_rho_T_cc_m}, we show the average QoE achieved by \textit{tailored federated position-only} predictor versus the assigned resources rate and SDoP. The results using \textit{trivial-motion} predictors are similar, hence are not provided for conciseness. We can observe that no matter how much the resources rate is assigned, which predictor is employed, the average QoE can always be improved with the increase of $\rho_s$. Besides, either larger value of $\rho_s$ or higher resources rate can support better QoE. For example, consider the point ``P". To achieve QoE = 94\%, one can increase $\rho_s$ by 0.2 or increase $R_{\mathrm{cc}}^*$ by 1.4. 

To explain why better QoE can be achieved with larger SDoP and visualize how much resources is traded for SDoP, we consider an example case in Fig. \ref{Fig:QoE, CC capability, and DoO v.s. DoP}. It is shown that a lager value of SDoP needs higher CC capability $C_{\mathrm{cc}}(t_{\mathrm{com}}^*, t_{\mathrm{cpt}}^*)$ but degrades average-DoO $\mathcal{D}(t_{\mathrm{obw}}^*)$, as expected. Since the degradation of average-DoO is smaller (from 76\% to 61\%), than the increase of the CC capability (from 0.18 to 1), the QoE is larger for a user with higher SDoP requirement. 

\begin{figure}[htbp]
	\vspace{-0.3cm}
	\centering
	\begin{minipage}[t]{0.85\linewidth}
		\includegraphics[width=1\textwidth]{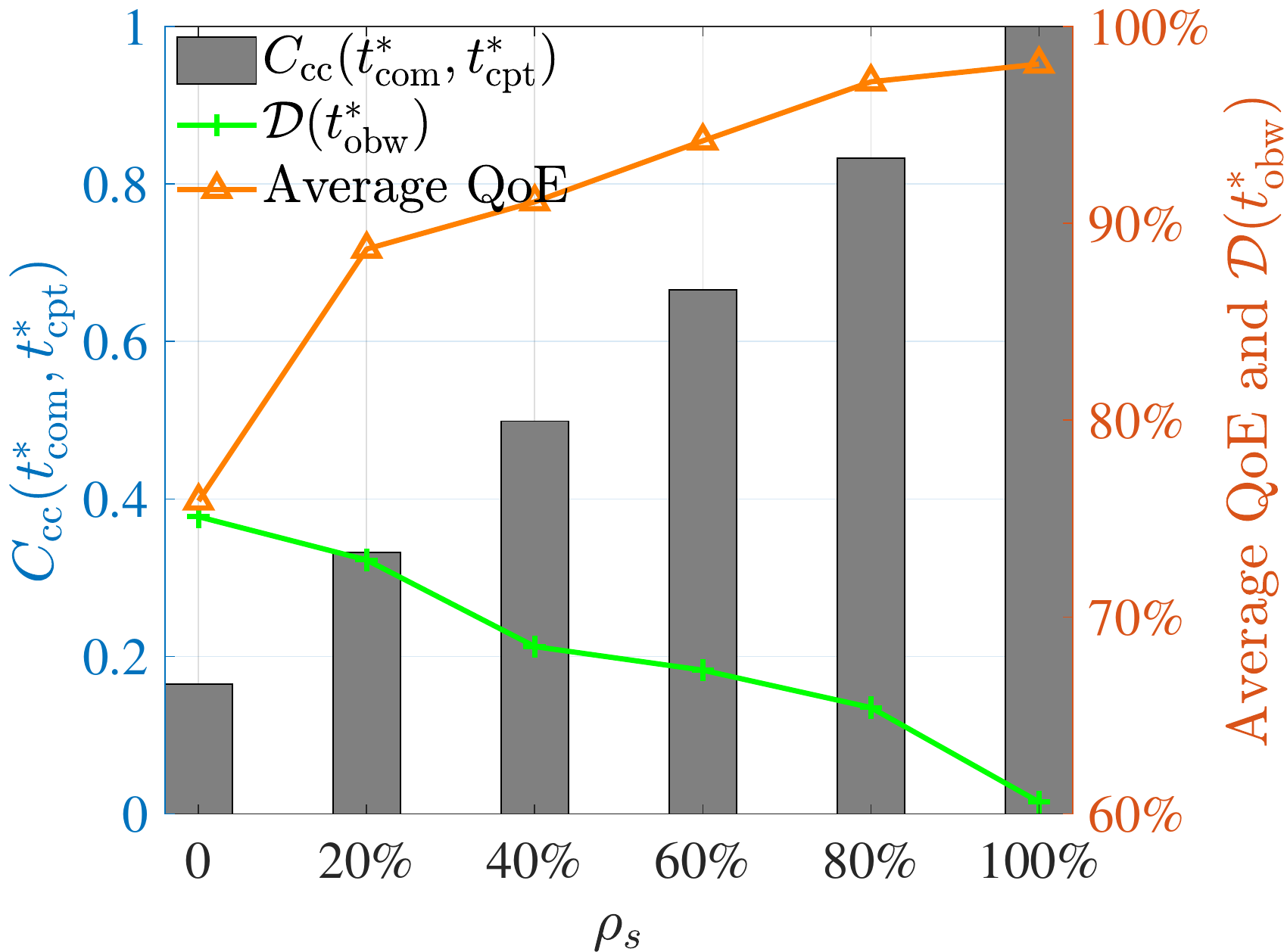}
	\end{minipage}
	\vspace{-0.2cm}
	\caption{Average QoE, CC capability, and average-DoO v.s. SDoP, $R_{\mathrm{cc}}^*=0.6$.}\label{Fig:QoE, CC capability, and DoO v.s. DoP}
	\vspace{-0.4cm}
\end{figure}


\section{Conclusion}
In this paper, we defined spatial degree of privacy to reflect the requirement for protecting the FoV of a user and investigated the impact of SDoP on proactive VR video streaming. When the configured resources for a user can satisfy the SDoP requirement of the user, we found from the optimized durations that a large value of SDoP degrades the performance of predicting tile requests due to requiring more resources. 
Simulation with the state-of-the-art predictors on a real dataset validated the analysis, and showed that by providing more resources to a user requiring larger SDoP, the QoE of the user can also be improved  although the prediction performance degrades.

\bibliographystyle{IEEEtran}
\bibliography{IEEEabrv,ref}

\end{document}